\documentclass[12pt]{emulateapj}

\usepackage{graphicx}

\newcommand{\ha}{H$\alpha$}
\newcommand{\solar}{\ifmmode_{\sun}\else$_{\sun}$\fi}
\newcommand{\HI}{H$\,${\sc i}}

\begin{document}

\title{Mass-to-Light versus Color Relations for Dwarf Irregular Galaxies}

\author{Kimberly A.\ Herrmann\altaffilmark{1}, Deidre A.\ Hunter\altaffilmark{2}, 
Hong-Xin Zhang\altaffilmark{3,4}, and Bruce G.\ Elmegreen\altaffilmark{5}}

\altaffiltext{1}{Pennsylvania State University-Mont Alto, 1 Campus Drive, Mont Alto, PA  17237, USA; kah259@psu.edu}
\altaffiltext{2}{Lowell Observatory, 1400 West Mars Hill Road, Flagstaff, AZ  86001, USA; dah@lowell.edu}
\altaffiltext{3}{Pontificia Universidad Cat\'{o}lica de Chile, Avenida Libertador Bernardo O Higgins 340, Santiago, Regi\'{o}n Metropolitana, Chile; hzhang@astro.puc.cl}
\altaffiltext{4}{National Astronomical Observatories, Chinese Academy of Sciences, Beijing 100012, China}
\altaffiltext{5}{IBM T. J. Watson Research Center, 1101 Kitchawan Road, Yorktown Heights, NY  10598, USA; bge@us.ibm.com}

\begin{abstract}
We have determined new relations between $UBV$ colors and mass-to-light ratios ($M/L$) for dwarf irregular (dIrr) galaxies, as well as for transformed $g^\prime - r^\prime$.  These $M/L$ to color relations (MLCRs) are based on stellar mass density profiles determined for 34 LITTLE THINGS dwarfs from spectral energy distribution fitting to multi-wavelength surface photometry in passbands from the FUV to the NIR.  These relations can be used to determine stellar masses in dIrr galaxies for situations where other determinations of stellar mass are not possible.  Our MLCRs are shallower than comparable MLCRs in the literature determined for spiral galaxies.  We divided our dwarf data into four metallicity bins and found indications of a steepening of the MLCR with increased oxygen abundance, perhaps due to more line blanketing occurring at higher metallicity.
\end{abstract}

\keywords{galaxies: dwarfs --- galaxies: irregular --- galaxies: structure --- galaxies: fundamental parameters}

\section{Introduction} \label{sec-intro}
Knowledge of the stellar mass in galaxies is important for a wide range of science problems. The best way to determine the stellar mass is through examination of the stellar populations, from either a star-by-star census or spectral energy distribution (SED) fitting of luminosities and colors based on stellar population synthesis (SPS) models. However, frequently, such data are not available, and we turn instead to a mass-to-light ratio, $M/L$, coupled with information on the luminosity or surface brightness, to derive a stellar mass or mass density.

Many studies have explored various calibrations of $M/L$ to color relations (MLCRs) for Johnson-Cousins optical bands (Bell \& de Jong 2001; Portinari et al.\,2004, hereafter P+04; McGaugh \& Schombert 2014, hereafter MS14), for Sloan Digital Sky Survey (SDSS) optical filters \citep{t+11,rc15}, and even some for both sets of filters (Bell et al.\,2003, hereafter B+03; Gallazzi \& Bell 2009; Zibetti et al.\,2009, hereafter Z+09; Into \& Portinari 2013, hereafter IP13).  In particular, MS14 tested the self-consistency of MLCRs by applying relations from various studies (B+03, P+04, Z+09, IP13) and various bands (specifically $M/L$ in $V$, $I$, $K$, and 3.6~$\mu$m all as functions of $B-V$) to estimate masses for a sample of disk galaxies spanning over 10 mag in luminosity. They found reasonable agreement between the four studies in $V$, but determined revised MLCRs for $I$, $K$, and 3.6~$\mu$m to improve self-consistency.

For a large sample of galaxies from the Two Micron All Sky Survey (2MASS) and the SDSS, B+03 used models to construct linear fits of $\log_{10} (M/L)$ and SDSS colors as well as $B-V$ and $B-R$. The fits take the form:
\begin{equation}
\log_{10} (M/L)_{\lambda} = a_{\lambda} + b_{\lambda} \times \rm{color}.
\end{equation}
From there, one can go on, for example, to determine the stellar mass density profile, $\Sigma (r)$, from the surface brightness, $\mu$, and color profiles using the following from \citet{btp08}:
\begin{equation}
\log_{10} \Sigma = \log_{10} (M/L)_{\lambda} - 0.4(\mu_{\lambda}-m_{\rm{abs},\odot,\lambda}) + 8.629,
\end{equation}

\noindent where $m_{\rm{abs},\odot,\lambda}$ is the absolute magnitude of the Sun at wavelength $\lambda$ and $\Sigma$ is measured in $M_{\odot}$~pc$^{-2}$.

However, a reliable relation between some color and the $M/L$ is essential, and the MLCRs in the literature have largely been determined from models appropriate for spirals. It is questionable if any of these linear fits is suitable for dwarf galaxies, with lower metallicities and potentially different star formation histories (SFHs) than spirals. Therefore, we have determined relations between $M/L$ and colors for dwarf irregular (dIrr) galaxies. These relations, presented here, are based on empirical SED fitting of multi-wavelength surface photometry of a sample of 34 dIrrs \citep{z+12}. In \S \ref{sec-data} we describe the data on which our new MLCRs are based. In \S \ref{sec-formula} we present our MLCRs between stellar $M/L$ from the SED fitting and $UBVg^\prime r^\prime$ colors and compare them to several MLCRs from the literature.  We explore metallicity effects in \S \ref{sec-Zbins} by breaking our data into four metallicity bins.

\begin{figure*}
\epsscale{0.9}
\plotone{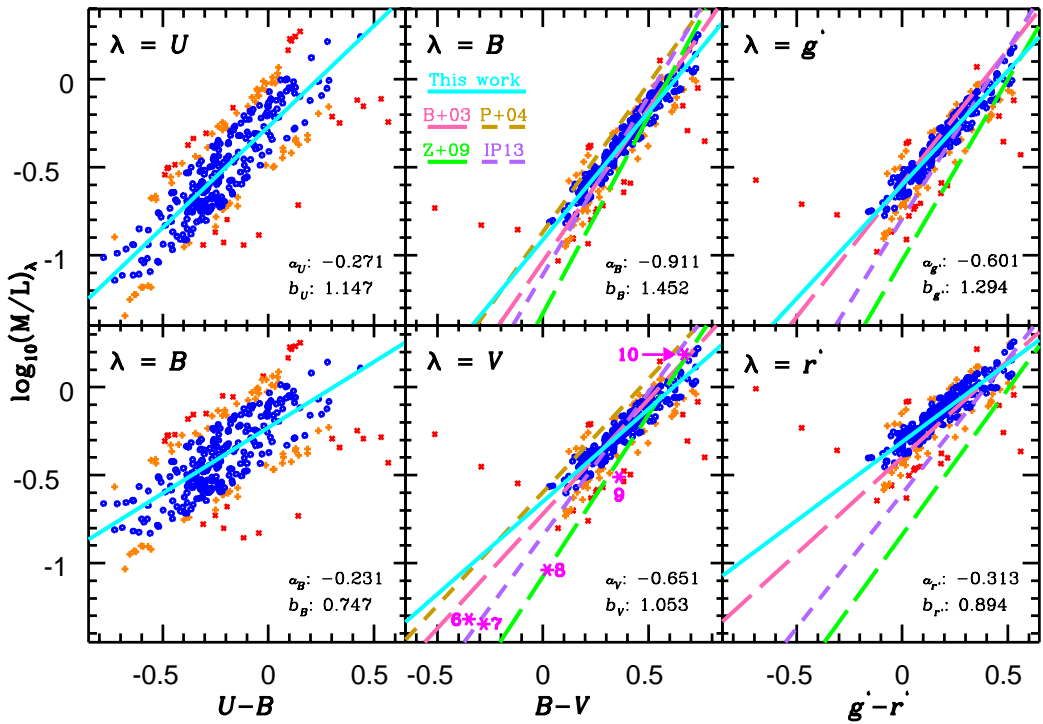}
\caption{New $M/L$ fits as a function of colors determined for dIrr galaxies. The small red x and orange + symbols show data with residuals larger than 0.3 and 0.2 dex, respectively, from the final fit in the two $U-B$ panels, but residuals larger than 0.2 and 0.1 dex, respectively, in the other four panels. Since the solid cyan fit was determined via iterative linear fits weighted by the inverse of the residuals, the blue data contributed more to the determination of each fit. The parameters of our final linear fits are provided following the equation $\log_{10}(M/L)_{\lambda} = a_{\lambda} + b_{\lambda} \times \rm{color}$. The additional colored lines show some MLCRs from the literature that have been shifted up or down as needed (see Table~1 for the parameters as well as its notes) to adjust to a \citet{c03} IMF.  Our MLCRs for dwarfs are slightly more shallow than those in the literature determined primarily for spirals. Lastly, the magenta asterisks show the $\log_{10}(M/L)_V$ for a single starburst population of 0.1-100 $M_{\odot}$ stars \citep{bc03} adjusted from its original Salpeter IMF. Each magenta number indicates the $\log_{10}$ age of the population in years.  Note that the $g'$ and $r'$ panels were derived from transformed $B$ and $V$ data using equations from \citet{SDSSfilters}.  \label{M_L} }
\end{figure*}

\section{Data} \label{sec-data}
The sample of galaxies is taken from LITTLE THINGS \citep[Local Irregulars That Trace Luminosity Extremes, The \HI\ Nearby Galaxy Survey,][]{lt12}. This is a multi-wavelength survey of 41 nearby ($<10.3$ Mpc) dIrr galaxies and blue compact dwarfs, which builds on the THINGS project \citep{walter08}, whose emphasis was on nearby spirals. The LITTLE THINGS galaxies were chosen to be gas-rich so they have the potential to form stars, although a few do not currently have \ha\ emission. They were also chosen to be fairly isolated, or at least not companions to a giant galaxy or obviously interacting with another system. The LITTLE THINGS data set includes {\it Galaxy Evolution Explorer} satellite \citep[{\it GALEX}; ][]{GALEX} images at FUV (1516 \AA) and NUV (2267 \AA) wavelengths, $UBV$ and some $JHK$ images from \citet{he06}, \ha\ images from \citet{he04}, and {\it Spitzer} Infrared Array Camera  \citep[IRAC]{irac04} 3.6 $\mu$m images. This yields 7-10 passbands for each galaxy from the FUV to the NIR.

\citet{z+12} used azimuthally averaged surface photometry in these passbands and performed a SED analysis of each annulus for a subsample of 34 of the LITTLE THINGS dwarfs.  They modeled the data with a library of four million different SFHs based on the unpublished 2007 version of GALAXEV (Bruzual \& Charlot 2003) stellar population models and allowing dust extinction, metallicity, and relative star formation rate among different age bins to vary uniformly among physically reasonable ranges. They used the stellar initial mass function (IMF) of \citet{c03} and took into account H$\alpha$ line emission, but not contamination from polycyclic aromatic hydrocarbon lines since nebular emission is arguably not important for the normal star-forming dIrr galaxies in the optical broadband images that were used. 

From the fits to the data, they derived stellar mass surface density distributions for the 34 dIrr galaxies. \citet{z+12} used two slightly different methods to determine the mass profile values at outer radii: (1) applying the $M/L$ from the most distant point where the SED analysis is applicable and (2) using the 3.6~$\mu$m light for some of the outermost points.  We used the latter mass profiles for the 34 dwarfs in our averaging analysis.

The SED fitting analysis yielded robust mass profiles as a function of radius in a large sample of dwarf galaxies. \citet{Paper1} and \citet{Paper2} have determined surface brightness and color profiles for these same galaxies, as well as 107 other dwarfs. With these data, we have the information we need to determine $M/L$ trends with $U-B$ and $B-V$ colors, as well as the transformed color $g'-r'$ (using $g'-r' = 0.98(B-V) - 0.19$, $g' = V + 0.54(B-V) - 0.07$, and $r' = V - 0.44(B-V) + 0.12$ from Smith et al.\,2002), that are appropriate specifically for dIrrs.

\begin{deluxetable*}{cccccccccccccc} \label{tab-fits}
\tablecaption{Fits as a Function of Color\tablenotemark{a}
\label{tab-fits}}
\tablewidth{0pt}
\tablehead{   &   & \multicolumn{4}{c}{This Work} & \multicolumn{2}{c}{B+03}
& \multicolumn{2}{c}{P+04} & \multicolumn{2}{c}{Z+09} & \multicolumn{2}{c}{IP13} \\
\colhead{Color} & \colhead{$\lambda$} & \colhead{$a_\lambda$} & \colhead{$\sigma(a_\lambda)$} & \colhead{$b_\lambda$} & \colhead{$\sigma(b_\lambda)$}
& \colhead{$a_\lambda$\tablenotemark{b}} & \colhead{$b_\lambda$} & \colhead{$a_\lambda$\tablenotemark{c}} & \colhead{$b_\lambda$}
& \colhead{$a_\lambda$} & \colhead{$b_\lambda$} & \colhead{$a_\lambda$\tablenotemark{c}} & \colhead{$b_\lambda$} }
\startdata
$U-B$ & $U$ & $-0.271$ & 0.081 & 1.147 & 0.260 & \dots & \dots & \dots & \dots & \dots & \dots & \dots & \dots  \\
$U-B$ & $B$ & $-0.231$ & 0.081 & 0.747 & 0.260 & \dots & \dots & \dots & \dots & \dots & \dots & \dots & \dots  \\
$B-V$ & $B$ & $-0.911$ & 0.157 & 1.452 & 0.393 & $-1.035$ & 1.737 & $-0.868$ & 1.690 & $-1.330$ & 2.237 & $-1.111$ & 2.027  \\
$B-V$ & $V$ & $-0.651$ & 0.157 & 1.053 & 0.393 & $-0.721$ & 1.305 & $-0.597$ & 1.290 & $-1.075$ & 1.837 & $-0.843$ & 1.627  \\
$g'-r'$ & $g'$ & $-0.601$ & 0.090 & 1.294 & 0.401 & $-0.592$ & 1.519 & \dots & \dots & $-1.030$ & 2.053 & $-0.794$ & 1.930  \\
$g'-r'$ & $r'$  & $-0.313$ & 0.090 & 0.894 & 0.401 & $-0.399$ & 1.097 & \dots & \dots & $-0.840$ & 1.654 & $-0.606$ & 1.530  \\
$V-3.6\mu$m & 3.6$\mu$m & $-0.175$ & 0.380 & $-0.149$ & 0.180 & \dots & \dots & \dots & \dots & \dots & \dots & \dots & \dots  \\
$B-V$ & 3.6$\mu$m & $-0.768$ & 0.148 & 0.776 & 0.367 & $-0.322$\tablenotemark{d} & $-0.007$\tablenotemark{d} & $-0.594$\tablenotemark{d} & 0.467\tablenotemark{d} & $-1.147$\tablenotemark{d} & 1.289\tablenotemark{d} & $-0.861$\tablenotemark{d} & 0.849\tablenotemark{d}
\enddata
\tablenotetext{a}{Fit parameters to $\log_{10}(M/L)_\lambda = a_\lambda + b_\lambda \times {\rm color}$}
\tablenotetext{b}{Reduced by 0.093 dex to adjust from ``diet'' Salpeter IMF to Chabrier IMF \citep{g+08} }
\tablenotetext{c}{Increased by 0.057 dex to adjust from Kroupa IMF to Chabrier IMF \citep{b+03,g+08} }
\tablenotetext{d}{From Table~7 of MS14}
\end{deluxetable*}

\begin{figure}
\epsscale{1.2}
\plotone{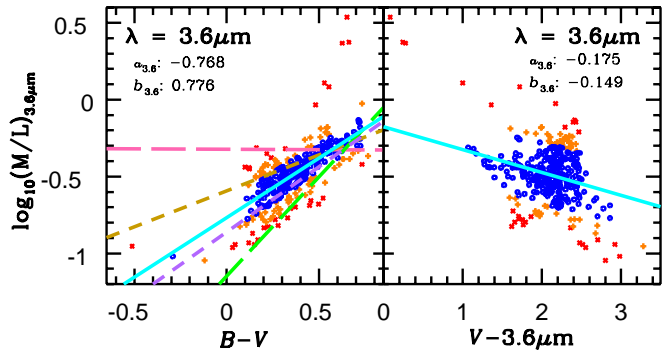}
\caption{New 3.6~$\mu$m $M/L$ fits as a function of colors determined for dIrr galaxies. The small red x and orange + symbols show data with residuals larger than 0.3 and 0.2 dex, respectively, in the $V - $3.6~$\mu$m panel, but residuals larger than 0.2 and 0.1 dex, respectively, in the $B-V$ panel. As in Figure~1, the solid cyan line shows our fit and the additional colored lines show some MLCRs from Table~7 of MS14 (see the legend in Figure~1 and see Table~1 for the parameters). $M/L$ in 3.6~$\mu$m is correlated well with $B-V$, but not strongly with $V - $3.6~$\mu$m. \label{new} }
\end{figure}

\section{Colors and mass-to-light ratios} \label{sec-formula}
The relations between $M/L$ and colors were determined via a minimizing $\chi^2$ fitting routine \citep{NumRec} iteratively reweighted to obtain a linear fit that was characteristic of as much of the data as possible, but not overly affected by outlying influential points.  The weights were 1.0 for $|\Delta y| <= \sigma$ where $\Delta y$ was the residual and $\sigma$ was the weighted standard deviation, both from each previous fit, and $\sigma/|\Delta y|$ for more discrepant data, following an L1 procedure for non-Gaussian residuals.  The $\chi^2$ fitting routine was iterated until the slope changed by $<10^{-8}$.  Figures~\ref{M_L} and \ref{new} show the optical and 3.6~$\mu$m data and fits, respectively, as well as the parameters of the final fits, which are also listed in columns 3 and 5 of Table~\ref{tab-fits}.

Stellar mass profiles are best calculated using the $B-V$ or $g'-r'$ relations; the $M/L$ ratio is not correlated as strongly with $U-B$ or $V-3.6 \mu$m as with $B-V$ or $g'-r'$, as seen by the larger scatter in the two $U-B$ panels of Figure~\ref{M_L} and the $V-3.6 \mu$m panel of Figure~\ref{new}.  This makes sense because redder passbands generally trace the mass better than bluer passbands.  FUV$-$NUV colors would not have been useful at all for estimating $M/L$, though $M/L$ in 3.6~$\mu$m is not as strongly correlated with $B-V$ as the $M/L$ in $B$ or $V$.

Since our relations (and those of Z+09) are based on a \citet{c03} stellar IMF, we have applied zero-point offsets to raise or lower MLCRs determined using other IMFs for a better comparison due to having one fewer variable between the different studies. \citet{bdj01} (and B+03) used a ``diet'' Salpeter IMF from 0.1 to 125 M\solar\ by modifying the $M/L$ by a factor of 0.7. This global factor was equivalent to reducing the contribution from low-mass stars, and was done to bring the absolute normalization of their $M/L$ in line with observational constraints of maximum disks determined from rotation curves of spiral galaxies. B+03 noted that $\log_{10} (M/L)$ must be reduced by 0.15 dex to convert from their ``diet'' Salpeter IMF to a \citet{k01} IMF.  \citet{g+08} specified a similar correction (-0.093 dex) to convert from the ``diet'' Salpeter IMF to a \citet{c03} IMF, which we applied to the MLCR of B+03. Since P+04 and IP13 both used Kroupa IMFs, we added 0.057 dex to adjust to a Chabrier IMF. To convert to a \citet{salpeter55} IMF, add a constant of +0.243 to the right side of Equation 1. For a \citet{k01} IMF, with a shallower IMF than Salpeter for low mass stars, subtract 0.057 dex. 

With all the literature MLCRs adjusted (as needed) to a Chabrier IMF, the B+03 optical relations almost overlap our linear fits and the data, except our optical fits for the MLCRs of dIrrs are consistently slightly shallower.  Actually, each cyan line (the dwarf best fit) in Figure~\ref{M_L} has the {\it shallowest} slope of all the colored lines in each panel, and correspondingly column~5 (dwarf $b_{\lambda}$ slopes) of Table~1 contains the smallest slope in each of the first six rows.  The Z+09 MLCRs provide a poor fit to our dIrr galaxy data for all five bands in the comparison.  The P+04 MLCRs are slightly high for all three $B-V$ panels whereas the IP13 MLCRs fit the dwarf data in the $B-V$ panels fairly well (except for being too steep) but fall below the dwarf data in both transformed $g'-r'$ panels.  The dwarf fit values for $\log_{10} (M/L_{r'})$ as a function of $g'-r'$ ($a_{r'} = -0.310$ and $b_{r'} = 0.890$) are similar to those used by \citet{btp08} for their spiral analysis ($a_{r'} = -0.306$ and $b_{r'} = 1.097$) although the latter $a_{r'}$ becomes $-0.399$ after being adjusted from the ``diet'' Salpeter IMF to a Chabrier IMF.

The magenta asterisks in the lower central ($B-V$, $V$) panel of Figure~\ref{M_L} show $\log_{10}(M/L)_V$ for a single starburst population of 0.1-100 $M_{\odot}$ stars \citep{bc03} reduced by 0.243~dex to adjust the data from their original Salpeter IMF to a Chabrier IMF.  Interestingly, the dwarf data fall somewhere between the 1~Gyr ($\ast$9) and 10~Gyr ($\ast$10) points since the typical luminosity-weighted age of dIrrs is $\sim$1~Gyr whereas spirals are generally older than dwarfs in terms of their luminosity-weighted ages.

\begin{figure*}
\epsscale{0.85}
\plottwo{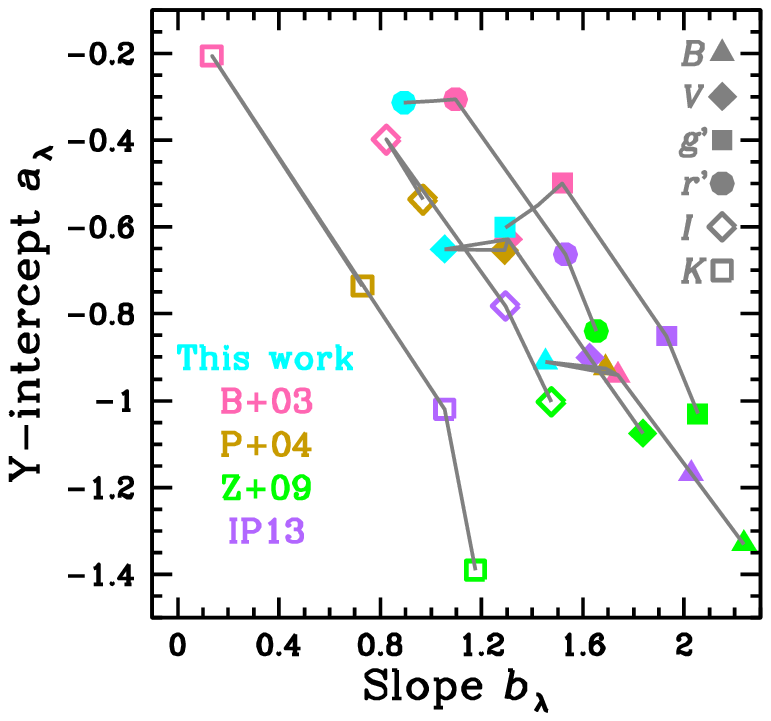}{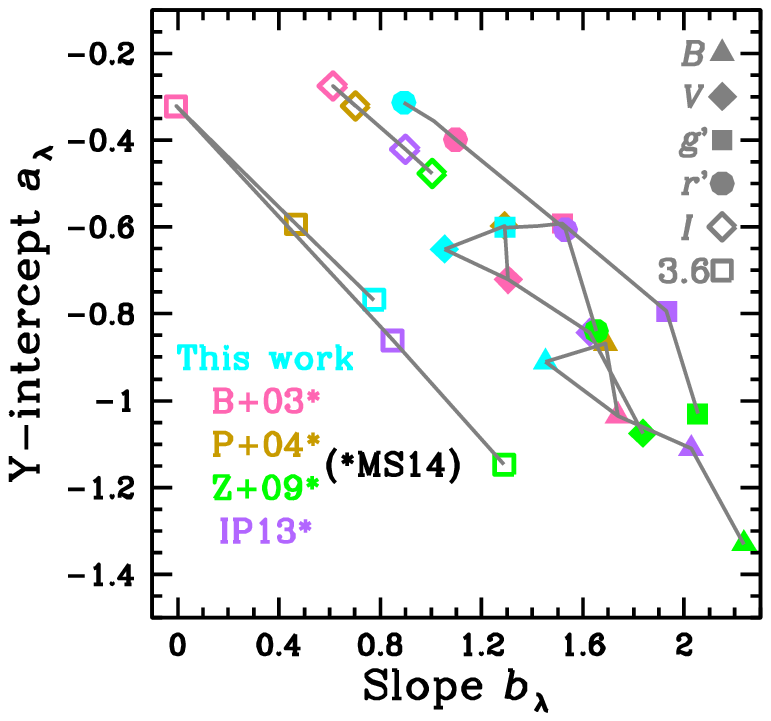}
\caption{MLCR linear fit parameters from various studies and for various bands where $\log_{10} (M/L)_{\lambda} = a_{\lambda} + b_{\lambda} \times \rm{color}$.  The color is $B-V$ for all the data points, except $g'-r'$ colors are used for $g'r'$.  In this work only, all $g'r'$ data have been transformed from $BV$ images. The left panel shows the original fit parameters from each study; the non-dwarf $V$, $I$, and $K$ data are directly from Table~2 of MS14 in terms of varying IMFs. The right panel shows the $BVg'r'$ $y$-intercepts adjusted as needed to a Chabrier IMF (the cyan and green data points are unchanged) and the revised $I$ and 3.6~$\mu$m fits from Table~7 of MS14. Note: (1) the general trend from $B$ (triangles) in the lower right through $r'$ (circles) and $I$ (open diamonds) at the upper middle and (2) the similar patterns between the various studies as well as between the various bands. \label{fit_params} }
\end{figure*}

Figure~\ref{fit_params} displays the fit parameters ($y$-intercepts, $a_{\lambda}$, and slopes, $b_{\lambda}$) of the MLCRs from this work as well as those from B+03, P+04, Z+09, and IP13.  The left panel shows the fit parameters from each study with the B+03 data based on the ``diet'' Salpeter IMF, the P+04 and IP13 data based on a Kroupa IMF, and the data from this work and Z+09 based on a Chabrier IMF.  The non-dwarf $V$, $I$, and $K$ data are directly from Table 2 of MS14. The right panel shows the $BVg'r'$ $y$-intercepts adjusted as needed to a Chabrier IMF (the cyan and green data points are unchanged) and the $I$ and 3.6~$\mu$m fits  from Table 7 of MS14, revised to be self-consistent with the $V$ band masses and corrected for $V-I$ as a second color term in addition to a 3.6~$\mu$m data point for dwarfs. The $BVIK$ and 3.6~$\mu$m fits are all with respect to $B-V$; that is, they are of the form: $\log_{10} (M/L)_{\lambda} = a_{\lambda} + b_{\lambda} (B-V)$ whereas the $g'r'$ fits are with respect to $g'-r'$.  In this work only, all $g'r'$ data have been transformed from $BV$ images. For completeness, we have included $IK$ MLCRs from the literature in Figure~\ref{fit_params} even though the LITTLE THINGS survey does not have enough observations in these bands for us to determine $IK$ MLCRs for dwarfs.

\begin{figure*}
\epsscale{0.88}
\plottwo{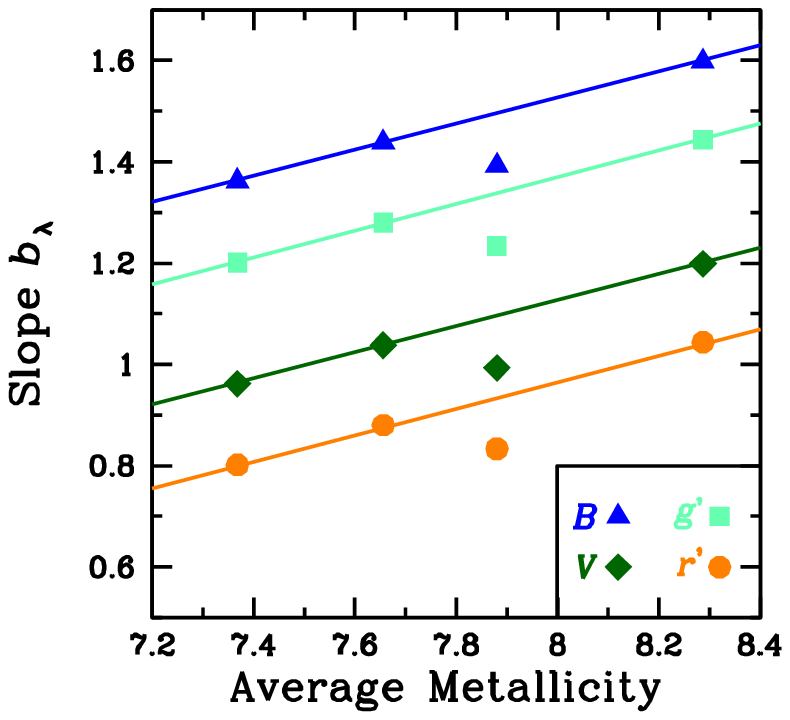}{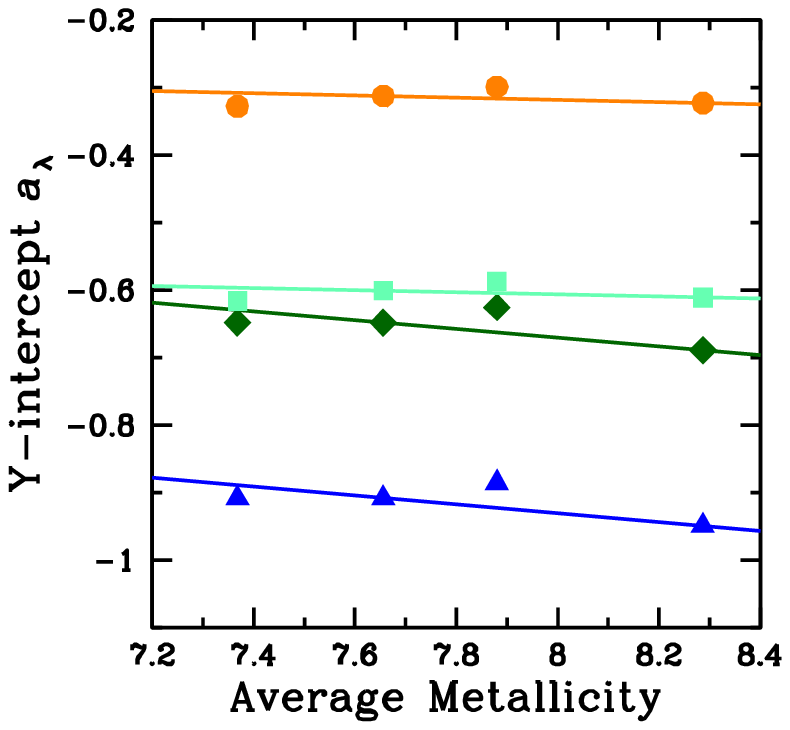}
\caption{MLCR linear fit parameters determined for four different metallicity bins where $\log_{10} (M/L)_{\lambda} = a_{\lambda} + b_{\lambda} \times \rm{color}$.  As before, the color is $B-V$ for the $\lambda = B$ and $\lambda = V$ points but $g'-r'$ for the $g'$ and $r'$ points.  We include the $g'r'$ data for future reference even though they are transformed from $BV$ observations.  The left panel shows a general steepening trend with higher metallicity whereas the right panel shows little dependance of the $y$-intercept on metallicity. The parameters for the linear fits shown in the left panel are listed in Table~\ref{tab-Z}. \label{Z_params} }
\end{figure*}

The optical bands (but not the $K$ or 3.6~$\mu$m data) in both panels display a fairly smooth progression from steep slopes and most negative $y$-intercepts in $B$ (triangles, lower right) up to more shallow slopes and less negative $y$-intercepts in $r'$ (circles) and $I$ (open diamonds, both in the upper middle).  However, in the right panel, the data points are more streamlined and the $I$-band points are all located in the upper middle instead of paralleling the $B$ to $r'$ points in the left panel.  Furthermore, within each optical band, the shape obtained from connecting the data from lower right to upper left follows roughly the same pattern from Z+09 (green), IP13 (lavender), B+03 (pink), this work (cyan), to P+04 (gold).  The differences between the various studies probably occur due to the variety of sample galaxy SFHs or assumed SFHs since SED masses and luminosities depend strongly on SFHs.

Why do the coefficients $a_{\lambda}$, the $y$-intercepts, and $b_{\lambda}$, the slopes, vary systematically with each other and with wavelength?  The fact that $a_{\lambda}$ decreases (the $M/L$ gets smaller) when $b_{\lambda}$ increases (the color dependence gets stronger) for bluer passbands means that younger regions with small $M/L$s have a greater dependence between $M/L$ and $B-V$ color than older regions.  This must occur since all the MLCR lines in Figure~1 have similar $M/L$ at the reddest colors.  That is, if we think of $M/L$ versus $B-V$ as pinned at a particular $M/L$ for a particular large $B-V$, and all the other curves varying as straight lines around that pinning point, then naturally the $y$-intercept $a_{\lambda}$ will get smaller as the slope $b_{\lambda}$ gets larger.  Physically this means that for red colors, the IMF and SFHs do not matter much because only old red stars remain and the $M/L$ and colors are universal for very old populations.  The variety of SED models and galaxy fits is generally due to different histories or assumptions about {\it young} stars.

\section{MLCRs and Metallicity} \label{sec-Zbins}

Since our MLCRs for dwarfs are shallower than the MLCRs in the literature and a lower metallicity is one characteristic that separates dwarfs from spirals, we broke our dwarf data into four bins based on oxygen abundance ($7.0 - 7.5$, $7.6 - 7.7$, $7.8 - 8.0$, and $8.1 - 8.7$) and fit the MLCR for each bin. Figure~\ref{Z_params} shows the results, including an apparent steepening of the MLCRs with higher metallicity (see Table~\ref{tab-Z} for the fit parameters) but little metallicity dependence of the $y$-intercepts, especially in $g'$ and $r'$.  Since the $g'$ and $r'$ analyses are not based on images but instead on transformations of $B$ and $V$, it is not surprising that the $g'$ and $r'$ fits have similar slopes to the $B$ and $V$ fits.  Unfortunately, the metallicities of the samples of spiral galaxies explored to determine existing MLCRs are not available for comparison.  However, for our slope trends from the left panel of Figure~\ref{Z_params} to match the B+03 MLCR slopes, the metallicity for the spiral sample would need to be somewhere between 8.5 and 8.8, which might be reasonable for a galaxy-wide average considering that the solar value is 8.69 \citep{a+09}.  Perhaps the shallower slope with lower metallicity is caused by less line blanketing occurring at lower metallicity \citep{h+15}.  The color changes with changing line blanketing (bluer for less line blanketing) {\it faster} than the brightness (brighter for less line blanketing) of a population.

\begin{deluxetable}{cccccc} \label{tab-Z}
\tablecaption{MLCR Slope Fits as a Function of Metallicity
\label{tab-Z}}
\tablewidth{0pt}
\tablehead{\colhead{Color} & \colhead{$\lambda$} & \colhead{Slope} & \colhead{$y$-intercept} & \colhead{$\sigma$(slope)} & \colhead{$\sigma (y$-intercept)}}
\startdata
$B-V$ & $B$ & 0.258 & $-0.537$ & 0.004 & 0.026 \\
$B-V$ & $V$ & 0.258 & $-0.937$ & 0.004 & 0.026 \\
$g'-r'$ & $g'$ & 0.264 & $-0.742$ & 0.004 & 0.029 \\
$g'-r'$ & $r'$  & 0.263 & $-1.139$ & 0.004 & 0.028 
\enddata
\end{deluxetable}

\section{Summary} \label{sec-summary}
We have used the stellar mass density radial profiles of \citet{z+12} produced from SED fitting to 7-10 passbands from the FUV to the NIR and the surface brightness and color profiles of \citet{Paper1,Paper2} to determine relations between $\log_{10}(M/L)$ and $U-B$ and $B-V$ colors as well as the transformed color $g'-r'$ given by \citet{SDSSfilters}. These MLCRs are specifically for dIrr type galaxies. The $B-V$ relationship in particular can be used to determine stellar masses in dIrrs in situations where SED fitting of multi-wavelength photometry or a census of individual stars is not available.  Our MLCRs are consistently shallower than those reported in the literature and the differences could be due at least partly to a steepening of MLCRs with metallicity, perhaps from more line blanketing occurring at higher metallicity.

\acknowledgments
We would like to thank the anonymous reviewer for helpful suggestions.  KAH acknowledges funding from Pennsylvania State-Mont Alto that enabled extended visits to Lowell Observatory where some of this work was carried out. DAH also appreciates funding from the Lowell Research Fund and is grateful to John and Meg Menke for funding for page charges. HZ acknowledges support from FONDECYT Postdoctoral Fellowship Project No. 3160538 and earlier from the Chinese Academy of Sciences (CAS) through a CAS-CONICYT Postdoctoral Fellowship administered by the CAS South America Centre for Astronomy (CASSACA) in Santiago, Chile.

\end{document}